\input phyzzx.tex
\tolerance=1000
\voffset=-0.0cm
\hoffset=0.7cm
\sequentialequations
\def\rl{\rightline}

\def\t1{{\tilde 1}}

\def\t{\theta}

\REF{\BEK}{J. Bekenstein, Lett. Nuov. Cimento {\bf 4} (1972) 737; Phys Rev. {\bf D7} (1973) 2333; Phys. Rev. {\bf D9} (1974) 3292.}
\REF{\HAW}{S. Hawking, Nature {\bf 248} (1974) 30; Comm. Math. Phys. {\bf 43} (1975) 199.}
\REF{\LEN}{L. Susskind, hep-th/9309145.}
\REF{\SBH}{E. Halyo, A. Rajaraman and L. Susskind, Phys. Lett. {\bf B392} (1997) 319, hep-th/9605112.}
\REF{\HRS}{E. Halyo, B. Kol, A. Rajaraman and L. Susskind, Phys. Lett. {\bf B401} (1997) 15, hep-th/9609075.}
\REF{\EDI}{E. Halyo, Int. Journ. Mod. Phys. {\bf A14} (1999) 3831, hep-th/9610068; Mod. Phys. Lett. {\bf A13} (1998), hep-th/9611175.}
\REF{\VEN}{T. Damour and G. Veneziano, Nucl. Phys. {\bf B568} (2000) 93, hep-th/9907030.}
\REF{\DAM}{T. Damour, gr-qc/0104080.}
\REF{\ULF}{U. Danielsson, A. Guijosa and M. Kruczenski, JHEP {\bf 0109} 011 (2001), hep-th/0106201.}
\REF{\KAU}{R. Kaul, Phys. Rev. {\bf D68} (2003) 024026, hep-th/0302170.}
\REF{\UNI}{E. Halyo, JHEP {\bf 0112} 005 (2001), hep-th/0108167.}
\REF{\DES}{E. Halyo, hep-th/0107169.}
\REF{\DAN}{N. Iizuka, D. Kabat, G. Lifschytz and D. Lowe, hep-th/0306209.}
\REF{\ATI}{J. J. Atick and E. Witten, Nucl. Phys. {\bf B310} (1988) 291.}
\REF{\KKS}{M. Karliner, I. Klebanov and L. Susskind, Int. Journ. Mod. Phys. {\bf A3} (1988) 1981.}
\REF{\HOL}{G. 't Hooft, gr-qc/9310026; L. Susskind, J. Math. Phys. {\bf 36} (1995) 6377, hep-th/9409089.}
\REF{\RAP}{R. Bousso, JHEP {\bf 9907} (1999) 004, hep-th/9905177; JHEP {\bf 9906} (1999) 028, hep-th/9906022; JHEP {\bf 0104} (2001) 035, hep-th/0012052.}
\REF{\STR}{L. Susskind, Journ. Math. Phys. {\bf 36} (1995) 36, hep-th/9409089.}
\REF{\POL}{G. Horowitz and J. Polchinski, Phys. Rev. {\bf D55} (1997) 6189, hep-th/9612146.}
\REF{\GIB}{G. Gibbons and S. Hawking, Phys. Rev. {\bf D15} (1977) 2738.}
\REF{\ADS}{E. Halyo, JHEP {\bf 0206} 012 (2002), hep-th/0201174.}
\REF{\HST}{G. Horowitz and A. Strominger, Nucl. Phys. {\bf B360} (1991) 197.}
\REF{\KLE}{I. R. Klebanov and A.A. Tseytlin, Nucl. Phys. {\bf B475} (1996) 164, hep-th/9604089.}
\REF{\BTZ}{M. Banados, C. Teitelboim and J. Zanelli, Phys. Rev. Lett. {\bf 69} (1992) 1849; M. Banados, M. Henneaux, C. Teitelboim and J. Zanelli, Phys.
Rev. {\bf D48} (1993) 1506.}
\REF{\BIT}{O. Bergman and C. Thorn, Phys. Rev. {\bf D52} (1995) 5980, hep-th/9506125; O. Bergman, hep-th/9606039.}
\REF{\MAT}{T. Banks, W. Fischler, S. Shenker and L. Susskind, Phys. Rev. {\bf D55} (1997) 5112, hep-th/9610043.}

\singlespace
\rl{hep-ph/0308166}
\rl{\today}
\pagenumber=0
\normalspace
\medskip
\bigskip
\titlestyle{\bf{Gravitational Entropy and String Bits on Stretched Horizons}}
\smallskip
\author{ Edi Halyo{\footnote*{e--mail address: vhalyo@stanford.edu}}}
\smallskip
\centerline{California Institute for Physics and Astrophysics}
\centerline{366 Cambridge St.}
\centerline{Palo Alto, CA 94306}
\smallskip
\vskip 2 cm
\titlestyle{\bf ABSTRACT}

We show that the entropy of Schwarzschild black holes in any dimension can be described by a gas of free string bits at the stretched horizon. 
The number of string bits is equal to the black hole entropy and energy dependent. For an asymptotic observer the bit gas is at the Hawking temperature. 
We show that the same description is also valid for de Sitter space--times in any dimension.

\singlespace
\vskip 0.5cm
\endpage
\normalspace

\centerline{\bf 1. Introduction}
\medskip

One of the most important and puzzling questions in quantum gravity is the origin of the Bekenstein--Hawking entropy of black holes[\BEK,\HAW]. It is well--known that the entropy in 
all cases is given by
(a quarter of) the area of the black hole horizon in Planck units. However, a microscopic explanation for this gravitational entropy due to horizons is not available. 
In particular, we do not know what the degrees of freedom counted by entropy are. These may be (related to) the fundamental degrees of freedom of quantum gravity.

A universal counting of black hole entropy in string theory would constitute an important step in this direction. Based on refs. [\LEN-\ULF], such a universal explanation 
for horizon entropy 
was given in ref. [\UNI]. In this framework, the black hole is described by a highly excited and long string on the stretched horizon. The string is at its Hagedorn temperature and has a 
rescaled tension due to
the large gravitational redshift. The entropy of the string gives the black hole entropy whereas the Hawking temperature is the redshifted Hagedorn temperature of the string.
This description is valid for all black holes (in all dimensions) with a near horizon space that is Rindler space. In addition, it applies 
to de Sitter space--times (in any dimension) which have the same property[\DES]. 

Recently, another (universal) description of black hole entropy was postulated. According to ref.[\DAN], the black hole entropy is due to a gas of noninteracting quasiparticles 
on the stretched horizon whose number 
is equal to the entropy. These quasiparticles have an average energy equal to the Hawking temperature. By definition, the number of quasiparticles depends on the temperature. 
This description may be improved in a number of ways. First, a gas with the above properties is not known; in ref. [\DAN] its existence was simply postulated. Second, it would be good
to obtain the same description in string theory and therefore base it on firmer ground. Finally, it would be better if this would be related to the universal counting of black 
hole entropy in terms of a long, highly excited string on the stretched horizon.

In this paper we show that, for a Schwarzschild black hole, the bits of the highly excited string on the stretched horizon have exactly the properties postulated 
for the quasiparticle gas. A highly excited
string has a length proportional to its entropy, $S \ell_s$. Therefore it can be considered as a collection of $S$ bits each with length $\ell_s$. Each bit has energy $1/\ell_s$
which seen from infinity becomes the Hawking temperature due to the gravitational redshift. This is not surprising since the string (and the string bits) are at the Hagedorn
temperature. We assume that the Hagedorn temperature is at or above the string binding energy at which there is a phase transition. Therefore we can consider the string bits to 
be free, e.g. the state at the Hagedorn temperature corresponds to the
deconfined phase of the string bits. Clearly, the properties of the string bit gas are very similar to those of the quasiparticle gas postulated in ref. [\DAN]. The only difference
between the two descriptions is the location of the stretched horizon; the quasiparticle gas is postulated to be a Planck (proper) length away from the event horizon whereas in the string
picture the (proper) distance is naturaly the string length. We show that the above 
reasoning also applies to de Sitter space--times (in any dimension) since the near horizon geometry is also Rindler space. 

The paper is organized as follows. In section 2 we review the description of Schwarzschild black hole entropy in terms of a highly excited string and argue that the string bits 
behave as a free gas of quasiparticles as in ref. [\DAN]. In section 3 we show the same for de Sitter space--times. Section 4 contains our conclusions and a discussion of our results.

\bigskip
\centerline{\bf 2. Black Hole Entropy and String Bits on the Stretched Horizon} 
\medskip

In this section, we first review the results of ref. [\SBH] which shows that the entropy of Schwarzschild black holes in any dimension can be described by a long, highly excited
string on the stretched horizon. The string has a rescaled tension due to the gravitational redshift and is at its Hagedorn temperature $T_{Hag} \sim 1/ \ell_s$. We then argue that 
this long string can be considered to be made of $S$ string bits each with length $\ell_s$ and energy $1/\ell_s$. We assume that the Hagedorn temperature $T_{Hag}$ is at or above the
the binding energy of the string bits. Therefore we can consider the bits to be in the deconfined phase of string theory. We find that there are $S$ 
noninteracting string bits on the stretched horizon where $S$ is the entropy of the black hole. For an asymptotic observer at infinity, the energy of a bit is the Hawking temperature
$T_H$ due to the gravitational redshift. This description is very similar to the one postulated in ref. [\DAN] in terms of free quasiparticles on the stretched horizon.

Consider the metric for the Schwarzschild black hole in $D$ dimensions
$$ds^2=-(1-{r_0 \over r})dt^2+(1-{r_0 \over r})^{-1}dr^2 +r^2 d \Omega_{D-2} \eqno(1)$$  
where 
$$r_0={{16 \pi G_D M} \over {(D-2) A_{D-2}}} \eqno(2)$$
with $G_D$ the $D$--dimensional Newton constant and $A_{D-2}=2 \pi^{(D-1)/2}/\Gamma({{D-1} \over 2})$.
The black hole entropy is given by
$$S_{BH}={M^{(D-2)/(D-3)} \over {4G_D}} A_{D-2}^{-1/(D-3)} \left({16 \pi G_D}\over {D-2} \right)^{(D-2)/(D-3)} \eqno(3)$$
whereas the Hawking temperature is
$$T_{H}={(D-2) \over {4(D-3)}} A_{D-2}^{-1/(D-3)} \left({{16 \pi } \over {(D-2) A_{D-2}}}\right)^{(D-2)/(D-3)} (G_D M)^{-1/(D-3)} \eqno(4)$$

Near the horizon, we define the coordinate $y=r-r_0^{1/(D-3)}$. The proper distance to the horizon is given by
$$R={2 \over {\sqrt{D-3}}} \sqrt y r_0^{1/2(D-3)} \eqno(5)$$
Then the coefficient of the $dt^2$ term becomes
$${1-{r_0 \over r}}={(D-3)^2 \over 2} R^2 \left({{16 \pi G_D M} \over {(D-2) A_{D-2}}} \right)^{-2/(D-3)} \eqno(6)$$
As a result, the dimesionless Rindler time is
$$\tau_R={(D-3) \over 2} \left({{16 \pi G_D M} \over {(D-2) A_{D-2}}}\right)^{-1/(D-3)} t \eqno(7)$$
The dimensionless Rindler energy is conjugate to $\tau$ and therefore $[E_R,\tau_R]=1$. Thus,
$$1={(D-3) \over 2} \left({{16 \pi G_D M} \over {(D-2) A_{D-2}}}\right)^{-1/(D-3)} [E_R,t] \eqno(8)$$
Since the black hole energy (mass) $M$ is conjugate to $t$
$$1={(D-3) \over 2} \left({{16 \pi G_D M} \over {(D-2) A_{D-2}}}\right)^{-1/(D-3)}{{\partial E_R} \over {\partial M}} \eqno(9)$$
This gives the Rindler energy
$$E_R=M^{(D-2)/(D-3)} {2 \over {D-2}} \left({{16 \pi G_D} \over {(D-2) A_{D-2}}} \right)^{1/(D-2)} \eqno(10)$$
The local energy of the string on the stretched horizon (located a proper length $\ell_s$ away from $r_0$) is $E_s \sim E_R/\ell_s$. This is much larger than the black hole mass $M$. 
Of course, seen from infinity the string
energy equals $M$ due to the gravitational redshift. Since for a string $E \sim \sqrt n/\ell_s$ we can identify the square root of the string oscillator number with the Rindler energy,
$\sqrt n \sim E_R$. Then the string entropy is $S \sim \sqrt n \sim E_R$. On the other hand, from eqs. (3) and (10) we see that
the entropy of the black hole is proportional to the Rindler energy, $S_{BH} \sim E_R$. Therefore the black hole entropy is described by the highly excited string on the stretched horizon.
The local temperature on the stretched horizon is $\sim 1/\ell_s$ which is the Hagedorn temperature of the string. For an asymptotic observer (after the redshift) this becomes 
the Hawking temperature in eq. (4).

Such a highly excited string is very long, with length $L \sim E \ell_s^2 \sim E_R \ell_s$. This can be shown as follows. Consider a string at level $n$. Its longitudinal length is[\SBH]
$$L=<\int_0^{2 \pi} d\sigma \sqrt{\partial _{\sigma}X_i \partial _{\sigma}X_i}>=2 \pi <\sqrt{\partial _{\sigma}X_i \partial _{\sigma}X_i}> \eqno(11)$$
A free string is a linear system and therefore
$$<\sqrt{\partial _{\sigma}X_i \partial _{\sigma}X_i}> \sim <\partial _{\sigma}X_i \partial _{\sigma}X_i>^{1/2}=(n \ell_s^2)^{1/2} \eqno(12)$$
where in the last step we used the virial theorem. Thus, we find $L \sim \sqrt n \ell_s$.

A highly excited, very long string can be considered to be made of $S \sim \sqrt n$ string bits each with length $\ell_s$. Since the string entropy is simply the number of bits 
we may conclude that each bit
carries one unit of entropy. As the string is excited, $n$ and therefore the number of bits increases and the string gets longer. The energy per bit is $E_{bit} \sim E/S
\sim 1/\ell_s$. This is not surprising since the string is at its Hagedorn temperature $T_{Hag} \sim 1/\ell_s$.
For a string below the Hageorn temperature, the bits are considered to be in the confined phase in which they are not free due to bit-bit interactions. However, there are reasons to 
believe that at the
Hagedorn temperature there is a phase transition to the deconfining phase in which the string breaks down into free bits[\ATI]. Since the string on the stretched horizon
is at $T_{Hag}$, we will consider it to be in the deconfined phase. 

As a result, we see that the black hole can be described by a free gas of $S$ string bits on the stretched horizon. The density of bits on the horizon is $\sim 1/G$ which
means one bit per unit horizon area. Each bit has energy $\sim 1/\ell_s$ which due to the redshift
becomes $T_H$ for an observer at infinity. Similarly the bit gas seems to be at the Hawking temperature for an asymptotic observer. The number of bits $S$ depends on
the temperature. This is exactly the description horizon entropy in terms of a gas of quasiparticles on the stretched horizon postulated in ref. [\DAN]. 
In our description, the string bits are the quasiparticles.
Note that (the string and) the string bits are at the Hagedorn temperature which is independent of energy. Thus the string bit gas has vanishing specific heat, $C_V=0$. Therefore
when energy is added to the gas of bits, the number of bits increases whereas the average bit energy and the temperature remain constant.
The negative specific
heat of the black hole is a result of the gravitational redshift which transforms the constant Hagedorn temperature into the energy dependent Hawking temperature.

Since the number of bits gives the black hole entropy, there is one bit per Planck area on the stretched horizon. In other words, the length of the string in string units is 
equal to the horizon area in Planck units. This shows that somehow the string must cover the area of the stretched horizon. In order to understand how this happens we need to find 
the dependence of the transverse size of the string on its
energy. Above we found that the longitudinal size of the string increases linearly with its energy. The transverse size increases much more slowly[\KKS]
$$R_{tr} \sim \ell_s log^{1/2}(E \ell_s) \eqno(13)$$
We see that the string length grows much faster than its transverse length. Therefore the string density or the number of bits per area will increase rapidly with energy.
At a certain energy the bit density per Planck area will be one. In string units, at that energy the bit density per area will be $1/g_s^2$. Beyond that energy, holography[\HOL,\RAP]
requires that the string bits behave as an incompressible fluid and the transverse size of the string grows like[\STR]
$$R_{tr}^{(D-2)} \sim \ell_P^{(D-2)} E \ell_s \eqno(14)$$

The energy at which the stringy behavior turns into the holographic one is given (approximately, up to a factor of $log^{1/(D-2)}(E \ell_s)$) by the string--black hole
correspondence principle[\POL]. This principle states that any black hole state corresponds to a state of a string where the two descriptions match when the black hole radius
is around the string length, $R \sim GM \sim \ell_s$. At this point the entropy of the string matches that of the black hole, i.e. $GM^2 \sim M \ell_s \sim \ell_s^2/G$. 
This is also the number of string bits on the stretched horizon and can be written as $S \sim 1/g_s^2$. This is precisely what we found above. We conclude that a string behaves
like a string if the number of its bits is $<1/g_s^2$ and as a black hole otherwise.

In our picture of the black hole, Hawking radiation is simply the emission of a bit from the bit gas on the stretched horizon. Since the bits are not interacting, we have on the 
stretched horizon
$${{dE} \over {dt}} \sim {\sqrt n E_{bit} \over \tau} \sim \sqrt n \ell_s^{-2} \eqno(15)$$ 
where $\tau \sim \ell_s^{-1}$ is the lifetime of a bit in the gas. For an asymptotic observer this becomes $dE/dt \sim \sqrt n T_H^2$ as expected.

As an example, consider a $D=4$ Schwarzschild black hole with mass $M$, entropy $S \sim GM^2$ and temperature $T_H \sim 1/GM$. This black hole is descibed by a string on the stretched
horizon with energy $GM^2/\ell_s>>M$ and entropy $GM^2$. The string is at its Hagedorn temperature $1/\ell_s$ at which it breaks down to its $S \sim \sqrt n \sim GM^2$ bits. Thus there is a
string bit gas on the stretched horizon. For an asymptotic observer, the temperature of the bit gas and the average energy per bit is $T_H \sim 1/GM$ due to the gravitational
redshift. The entropy of the gas is a number that does not redshift and it is given by the number of bits $S \sim GM^2$. We see that the number of bits is energy dependent.

\bigskip
\centerline{\bf 3. de Sitter Entropy and String Bits on the Stretched Horizon}
\medskip


We expect the above results to hold also for $D$-dimensional de Sitter space--times since the near horizon geometry of de Sitter space is Rindler space just like a Schwarzschild black hole.
For completeness we show this below which is review of ref. [\DES].

An inertial observer in de Sitter space--time sees a cosmological horizon. 
The static metric which describes the ($D$--dimensional) region inside the horizon is given by
$$ds^2=-\left(1-{r^2 \over L^2} \right)dt^2+\left(1-{r^2 \over L^2} \right)^{-1} dr^2+r^2 d\Omega_{D-2}^2 \eqno(16)$$
where $0 \leq r \leq L$ and the horizon is at $r=L$. 
The entropy of de Sitter space--time is given by the horizon area in Planck units[\GIB]
$$S_{dS}={A \over {4G_N}}={L^{D-2} A_{D-2} \over 4 G_D} \eqno(17)$$
where $A_{D-2}=2 \pi^{(D-1)/2}/\Gamma((D-1)/2)$ is the area of the $D-2$ dimensional unit sphere.
The temperature of de Sitter space--time (in any dimension) is
$$T_{dS}={1 \over {2 \pi L}} \eqno(18)$$

The near horizon geometry of de--Sitter space is given by Rindler space. Near the horizon taking $r=L-y$ with $y<<L$ we find that the metric becomes
$$ds^2= -\left(2y \over L \right)dt^2+\left(2y \over L \right)^{-1} dy^2+L^2 d\Omega_{D-2}^2 \eqno(19)$$
The proper distance to the horizon is 
$$R=\sqrt{L\over 8}\sqrt{y} \eqno(20)$$
Then the $t-R$ part of the metric becomes
$$ds^2=-\left(16 R^2 \over L^2 \right)dt^2+dR^2+\ldots \eqno(21)$$
which is the Rindler metric. The dimensionless Rindler time is 
$$\tau_R={4 \over L} t \eqno(22)$$
The dimensionless Rindler energy $E_R$ is conjugate to the Rindler time, i.e. $[E_R,\tau_R]=1=(4/L)[E_R,t]$ and therefore
$$1={4 \over L} {\partial E_R \over \partial E} \eqno(23)$$
where we used the fact that the energy $E$ is conjugate to time $t$. 
Eq. (23) gives the Rindler energy for de Sitter space
$$E_R \sim{L^{D-2} \over  G_D} \eqno(24)$$

Again, consider a string at the stretched horizon. The energy of this string is $E \sim E_R/\ell_s$ and
therefore,
as above we identify the Rindler energy $E_R$ with the square root of oscillator number $\sqrt{n}$ of the string. As a result,
$S \sim E_R \sim S_{dS}$.

Just as in the black hole case, we find that the entropy of de Sitter space--time in any dimension is given by the entropy of a string near the horizon with oscillator number $n \sim E_R^2$. 
This is a very long string of length $\sim E_R \ell_s$. Again, we assume that the string is made of $E_R$ bits each of length $\ell_s$ carrying one bit of information. The string bits are
the fundamental degrees of freedom on the horizon with a density of one per Planck area. Since the string is at its Hagedorn temperature $1/\ell_s$ we assume that it is in the
deconfined phase in which the bits are free.
We find that de Sitter entropy can be described by a gas of $E_R$ free string bits on the stretched horizon. The temperature of the gas and the average energy per bit is $1/\ell_s$. 
The entropy is the number of bits $E_R$.

On the stretched horizon, the energy of the string is $E \sim L^{(D-2)}/ (\ell_s G_D)$. However, 
the energy seen by an observer at the center of de Sitter space (at $r=0$) is given by 
$$E_{dS} \sim {L^{D-3} \over G_D} \sim {\sqrt{n \over T}} \sim {E_R \over \sqrt{T}}  \eqno(25)$$
where we identified the total energy seen at $r=0$ with the energy of the string.
We see that the tension of
the string is renormalized due to the large redshift to $T  \sim 1/L^2$. Thus, for an observer at $r=0$ the temperature of
the string becomes $\sim {1/ L}$
which is the temperature of de Sitter space--time in any dimension. 
We conclude that, for an observer at $r=0$, the thermodynamics of de Sitter space--time can be described by a gas of $E_R$ string bits at temperature $T_{dS}$. As before, the entropy
is the number of bits $E_R$ which is energy dependent.

\bigskip
\centerline{\bf 4. Conclusions and Discussion}
\medskip

In this paper, we showed that the entropy of Schwarzschild black holes and de Sitter space--times in any dimension can be obtained by that of a gas of string bits that lives on the
stretched horizon. The number of (free) string bits is equal to the horizon entropy. On the stretched horizon, the bits are the Hagedorn temperature, $1/\ell_s$ which is also 
the energy per bit. For the black hole (de Sitter space--time) case an asymptotic observer at infinity (at $r=0$) sees the bit gas at the Hawking (de Sitter) temperature due to the 
gravitational redshift. This result is based on the fact that in both cases the entropy can be described by that of a long, highly excited string on the stretched horizon with a
rescaled tension. This in turn is due to the fact that in both cases the near horizon geometry is Rindler space. 

Therefore, it seems that the fundamental degrees of freedom on the stretched horizon which count black hole or cosmological entropy are string bits. The long string on the stretched
horizon has exactly the right number of bits to give the correct entropy. Apparently, this string has also a large enough transverse size to cover the horizon so that
by the time the bits are deconfined there is one bit per Planck area. The mechanism that leads to this is not clear and probably requires a much better understanding of the physics
of string bits. Note that the string bit gas has very unique properties on the stretched horizon. The temperature (and the energy per bit) does not depend on the energy but the
number of bits increases with it. Therefore, the gas has a vanishing specific heat. For an asymptotic observer the gas seems to have an energy dependent temperature but a negative
spacific heat. In the above picture, these well--known results are the results of the gravitational redshift.

The property common to both Schwarzschild black holes and de Sitter space--times is the fact that in both cases the near horizon geometry is Rindler space. However, there are
many other cases with the same property such as asymptotically de Sitter space--times[\ADS], near--extreme (black) D--branes[\HST,\KLE] and the BTZ black hole[\BTZ]. 
In fact, exactly this property was 
used in ref. [\UNI] to show that the entropy in all these cases can be described by a string on the stretched horizon. Thus, it is clear that our results can be easily generalized to these
cases. It seems that the only cases in which gravitational entropy cannot be described by a long string or a gas of string bits on the stretched horizon are $D=4,5$ 
Reissner--Nordstrom black holes. These black holes have more than two charges and nonvanishing extreme entropies which cannot be (at least naively) described by strings. 

The string bit description of the horizon entropy coincides with that in terms of free quasipaticles postulated in ref. [\DAN]. As mentioned in the introduction, the only
difference between the two descriptions is the location of the stretched horizon which is a proper distance $\ell_s$ ($\ell_P$) away from the event horizon in the bit (quasiparticle) picture.
These distances are the natural ones in string theory (for bits) and semi--classical gravity (for quasiparticles). Note that for a small string coupling the difference between 
$\ell_s$ and $\ell_P$ can be very large. It is surprising that the two descriptions match so well even though the location of the stretched horizon is different.

The relation between the string bits on the stretched horizon and the those in the literature is not clear. String bits generally are introduced in order to describe
(nonperturbative) string theory on a (one--dimensional) lattice[\STR,\BIT]. The string bit models are described in the light--cone gauge which assumes that the string has a very 
high momentum in some
direction which is taken to be compact. The bits describe one unit of momentum in the compact direction whereas the total longitudinal momentum gives the number of bits.
In our case, there is no such compact direction and therefore no quantum of longitudinal momentum. In addition, the string is stationary on the stretched horizon. On the other hand,
for an asymptotic
observer the string seems to be boosted continuously as time passes with no limit which does not correspond to a high but fixed momentum. 
Another difference is that in our case the number of bits (or the length of the string) is 
given by the string energy and not its longitudinal momentum. One may speculate that the compact direction is the 11th direction of M--theory. Then the bit picture makes more sense
since it becomes the decription of M(atrix) theory[\MAT]. However, such a correspondence would imply that the string energy somehow describes momentum in the 11th direction which is
hard to justify.
It is amusing that both in our description and in M(atrix) theory the black hole entropy is given by the number of bits even though
there are many differences between the two pictures and types of bits. Clearly, the resolution of the above issues requires a better understanding of string bit physics and/or
the description of strings above the Hagedorn temperature.



\vfill

\refout

\end
\bye